\documentclass{emulateapj}
\usepackage{color,soul}
\pdfoutput=1
\date{\today}

\def\lya{Ly$\alpha$ }
\def\Msun{M_\odot}
\def\hMsun{h^{-1}M_\odot}
\def\kpc{{\rm kpc}}
\def\Mpc{{\rm Mpc}}
\def\hMpc{h^{-1}{\rm Mpc}}

\def\VhMpc{h^{-3}{\rm Mpc}^3}

\def\kms{{\rm km\, s^{-1}}}
\def\Rvir{R_{\rm vir}}
\def\SBunit{{\rm erg\, s^{-1}\, cm^{-2}\, arcsec^{-2}}}
\def\UVunit{{\rm erg\, s^{-1}\, cm^{-2}\, Hz^{-1}\, arcsec^{-2}}}

\slugcomment{{\sc Accepted to ApJ: April 9, 2015} }

\begin{document}
\bibliographystyle{apj}

\title{On the Diffuse Lyman-alpha Halo Around Lyman-alpha Emitting Galaxies}

\author{
Ethan Lake$^1$,
Zheng Zheng$^1$,
Renyue Cen$^2$,
Raphael Sadoun$^1$,
Rieko Momose$^3$,
and
Masami Ouchi$^{3,4}$
}
\altaffiltext{1}{Department of Physics and Astronomy, University of Utah,
 115 South 1400 East, Salt Lake City, UT 84112, USA; ethan.lake@utah.edu, zhengzheng@astro.utah.edu}
\altaffiltext{2}{Princeton University Observatory, Princeton, NJ 08544, USA}
\altaffiltext{3}{Institute for Cosmic Ray Research, The University of Tokyo, 5-1-5 Kashiwanoha, Kashiwa, Chiba 277-8583, Japan}
\altaffiltext{4}{Kavli Institute for the Physics and Mathematics of the Universe (WPI), The University of Tokyo, 5-1-5 Kashiwanoha, Kashiwa, Chiba 277-8583, Japan}

\begin{abstract}

\lya photons scattered by neutral hydrogen atoms in the circumgalactic media or
produced in the halos of star-forming galaxies are expected to lead to extended
\lya emission around galaxies. Such low surface brightness \lya halos (LAHs)
have been detected by stacking \lya images of high-redshift star-forming
galaxies. We study the origin of LAHs by performing radiative transfer 
modeling of nine $z=3.1$ Lyman-Alpha Emitters (LAEs) in a high resolution 
hydrodynamic cosmological galaxy formation simulation. We develop a method of computing the 
mean \lya surface brightness profile of each LAE by effectively integrating 
over many different observing directions. Without adjusting any parameters, 
our model yields an average \lya surface brightness profile in remarkable 
agreement with observations. We find that observed LAHs cannot be accounted 
for solely by photons originating from the central LAE and scattered to large 
radii by hydrogen atoms in the circumgalactic gas. Instead, \lya emission 
from regions in the outer halo is primarily responsible for producing the 
extended LAHs seen in observations, which potentially includes both 
star-forming and cooling radiation. With the limit on the star formation
contribution set by the ultra-violet (UV) halo measurement, we find
that cooling radiation can play an important role in forming the extended 
LAHs. We discuss the implications and caveats of such a picture.
 
\end{abstract}


\section{Introduction}

The \lya line is an important cosmological tool for studying star-forming galaxies in the young universe, and has been found to aid in the detection of high redshift galaxies \citep{Rhoads:2003aa, Ouchi:2007aa, Gawiser:2007aa, Guaita:2009aa}. As ionizing photons are emitted from young 
stars, they ionize neutral hydrogen in the surrounding interstellar medium, and are likely to be re-emitted 
as \lya photons following recombination \citep{Partridge:1967aa}. After they escape the 
interstellar medium surrounding their parent stars, they are predicted to undergo resonant scattering with 
neutral hydrogen gas in the surrounding medium as a result of the radiative transfer process, diffusing 
out both spatially and in frequency \citep{Zheng:2011ab}. As such, extended halos of neutral hydrogen 
around these Ly$\alpha$ emitters (LAEs) are predicted to be illuminated by scattered Ly$\alpha$ 
photons. 

Many theoretical studies have predicted the existence of these so-called 
Ly$\alpha$ halos (LAHs) around high redshift galaxies 
\citep[e.g.,][]{Tasitsiomi:2006aa,Laursen:2007,Laursen2009,Dijkstra:2009,
Barnes:2010,Barnes2011,Zheng:2010aa, Zheng:2011ab}.
While these LAHs are predicted to generally be too faint to be detected on an 
individual basis at $z$ $\geq 2$, their presence can be revealed by stacking tens 
to hundreds of narrow band images of high redshift LAEs \citep{Fynbo:2003aa,  
Steidel:2011aa, Zheng:2011ab, Matsuda:2012aa, Momose:2014aa}. Although 
observationally there seems to be a consensus in favor of their existence,
there are also reports of null detections of LAHs.
\citet{Feldmeier:2013aa} find marginal evidence and no evidence of LAHs for
$z\sim 3.1$ and $z\sim 2.1$ LAEs, while \cite{Jiang:2013aa} find evidence of 
LAHs based on results using stacked images of LAEs at redshifts of 5.7 and 6.6. 
Such contradicting results may be caused by small number statistics or unknown
systematics \citep{Momose:2014aa}.

The shape and size of LAHs can yield insights into the spatial distribution 
and kinematic properties of the circumgalactic and intergalactic medium 
surrounding LAEs \citep{Zheng:2011ab}. The shape can also be used to constrain
cosmic reionization, with reionization leading to steeper surface brightness 
profiles \citep{Jeeson-Daniel:2012aa}. Detailed theoretical studies of LAHs 
can help in understanding their origin and properties. 

The aim of this paper is to apply a Monte Carlo radiative transfer code 
\citep{Zheng:2002aa} to study diffuse Ly$\alpha$ halos surrounding 
$z\sim 3.1$ star-forming galaxies in a high-resolution galaxy formation 
simulation. By comparing with observational data, we hope to gain insight 
about the origin and composition of these diffuse LAHs.

\begin{figure*}
\epsscale{0.93}
\plotone{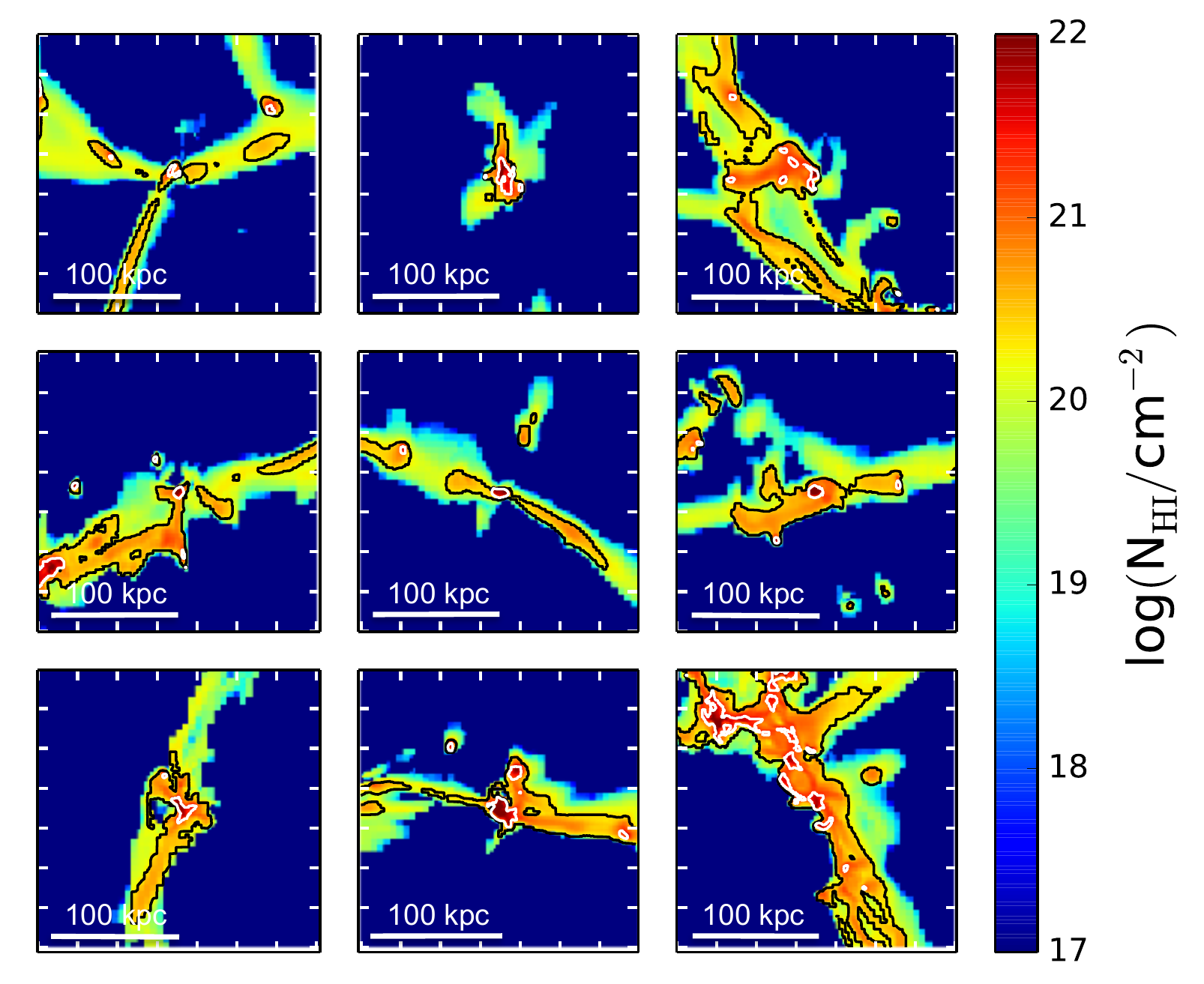}
\caption{
 \label{fig:nhimage}
Neutral hydrogen column density maps for the model LAEs in our analysis. 
Each image is 224~\kpc\ (physical) on a side. The column density is computed
by integrating over the whole box along the line of sight (224~\kpc\ physical).
The black contours are drawn
at $10^{20.3}$cm$^{-2}$, within which are regions corresponding to DLAs.
The white contours are drawn at $10^{21.3}$cm$^{-2}$. Extended filamentary 
structures of neutral hydrogen are seen, which are connected to the 
extended \lya emission discussed in this paper.
}
\end{figure*}

This paper is divided into several sections. In Section~2, we describe the 
modeling method and the construction of the average \lya surface brightness 
profile for each model LAE. Our main analyses and results are presented in 
Section~3, with comparisons to observations and a discussion of possible
constraints imposed by the profile in the UV band. 
Finally, we summarize our results and discuss the implications in 
Section~4.

\section{Model}

\subsection{\lya Radiative Transfer Modeling of Simulated Star-forming Galaxies}

\begin{figure*}
\epsscale{0.93}
\plotone{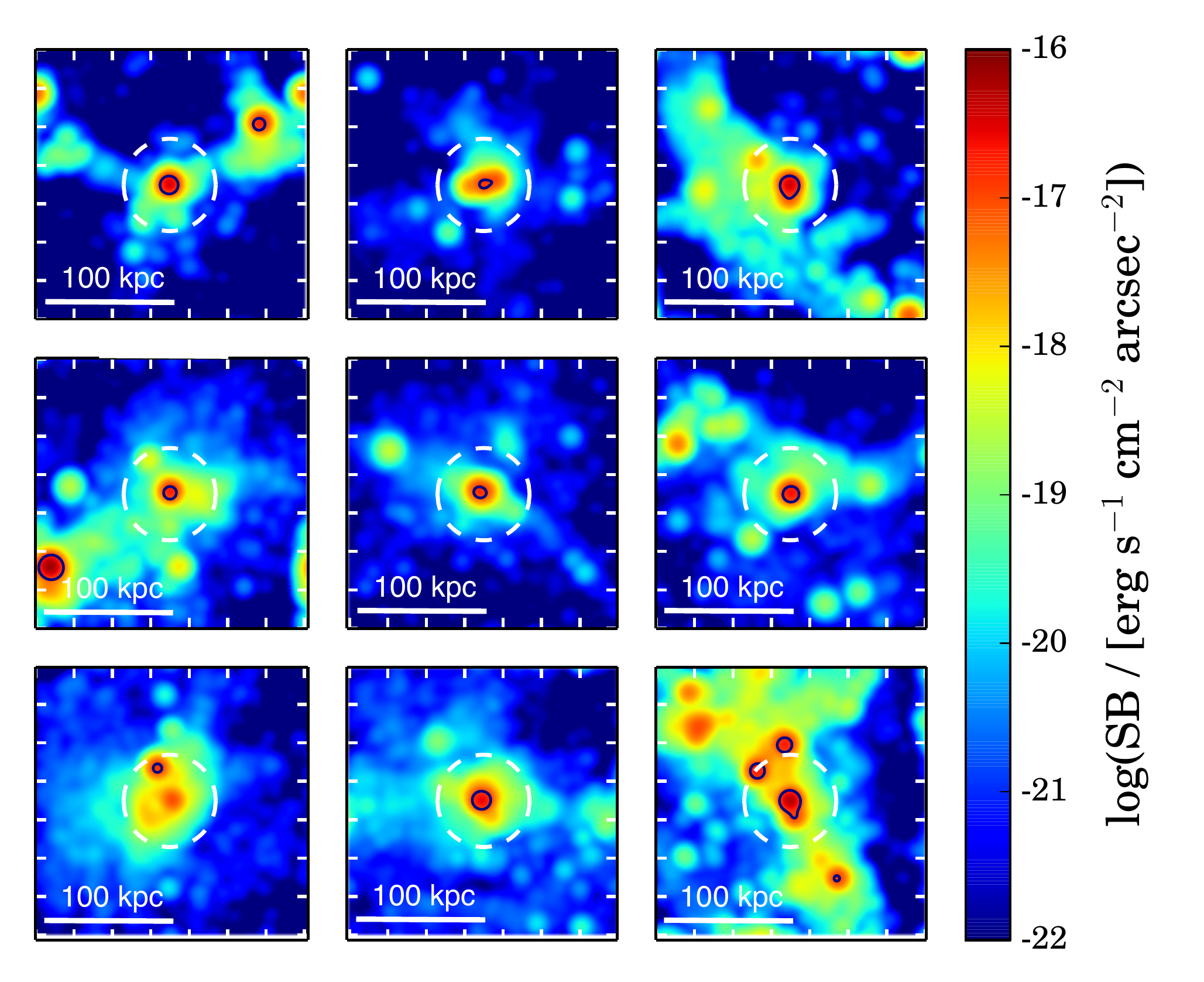}
\caption[]{
\label{fig:image}
\lya surface brightness images for all nine model LAEs in our analysis. Each 
image is 224~\kpc\ (physical) on a side and has been smoothed by a 2D Gaussian 
kernel with a FWHM of 1.32\arcsec\ to match the observation setup. 
Isophotal contours are drawn in black at the limit of observational 
detection, $10^{-17}\SBunit$. The dashed circle in each panel has a radius of
40~\kpc, beyond which the systematic effects in the image stacking analysis 
in \citet{Momose:2014aa} start to become important.  
The images show a rich diversity in structure at 
surface brightness levels below the detection threshold, which contributes to 
the extended LAHs.
}
\end{figure*}

Our \lya radiative transfer modeling of simulated star-forming galaxies is 
based on a cosmological simulation with the adaptive mesh refinement (AMR)
Eulerian hydro code Enzo \citep{Bryan:2000aa,Joung:2009aa}, as detailed in
\citet{Cen:2012aa} and \citet{Cen:2013aa}. 
In brief, a region of comoving size $21\times 24 \times 
20 \VhMpc$ in a low-resolution simulation (with a box size of 120$\hMpc$ 
comoving on a side) is chosen to be resimulated at high resolution. The 
resimulation has a dark matter particle mass of $1.3\times 10^7\hMsun$ and
the mesh refinement ensures a spatial resolution better than 
$111 h^{-1}{\rm pc}$ (physical). The resimulation includes an ionizing 
UV background and the self-shielding of the gas, 
metallicity-dependent radiative cooling,
molecular hydrogen formation, star formation, and supernova feedback. The mass
of a star particle is typically $\sim 10^6\Msun$. The simulation assumes
a spatially flat $\Lambda$CDM model with the following cosmological 
parameters: $\Omega_m=0.28$, $\Omega_b=0.046$, $H_0=100 h\,\kms\Mpc^{-1}$ with 
$h=0.70$, $\sigma_8=0.82$, and $n_s=0.96$.

The simulation has been used to study the kinematic properties traced by 
unsaturated metal lines in damped \lya systems, which is in good
agreement with observations \citep{Cen:2012aa}. The simulation has also
been applied to study the partition of stellar light into optical and 
infrared light \citep{Cen:2011aa}. In \citet{Cen:2013aa}, a model of
\lya blobs (LABs) is developed based on \lya radiative transfer modeling
of the simulated star-forming galaxies in massive halos, and the
observed relation between \lya luminosity and LAB size and LAB luminosity
function at $z\sim 3.1$ have been successfully reproduced. In this paper, 
we select from the simulation 9 $z=3.1$ star-forming galaxies in halos of 
mass $10^{11.5}\Msun$ to study the properties of LAHs associated with them. 
The value of $10^{11.5}\Msun$ is chosen as a starting point for our analysis, 
and is within current constraints on LAE halo mass of $10^{11\pm1}\Msun$ 
\citep{Ouchi:2010aa}. The mean stellar mass of these nine galaxies is about
$2.9\times 10^{10}\Msun$.

We implement a Monte Carlo code developed by \citet{Zheng:2002aa} for the
\lya radiative transfer calculation in extended neutral hydrogen distributions
surrounding our simulated LAEs. This code has been applied to study
LAEs and LABs \citep[e.g.,][]{Zheng:2010aa,Zheng:2011aa,Zheng:2011ab,
Cen:2013aa,Zheng:2014aa}. 
For each galaxy, we store the relevant quantities from the simulation in a 
uniform cubic grid of $4\Rvir$ on a side, with cell size 319 pc 
(physical, corresponding to 0.04\arcsec). Here $\Rvir$ is the virial 
radius of the host halo, which is on average $\sim 56~\kpc $ (physical)
for the nine $10^{11.5}\Msun$ halos we consider. The quantities
include the \lya luminosity, neutral hydrogen density, temperature, and 
velocity. 
The \lya luminosity is separated into star-formation and cooling contributions.
The \lya luminosity from star formation is computed as $L_{\rm Ly\alpha} 
= 10^{42}[{\rm SFR}/(\Msun{\rm yr}^{-1})] {\rm erg\, s^{-1}}$ 
\citep{Furlanetto:2005aa}, where SFR is the star formation rate in the cell.
The \lya luminosity from cooling radiation is computed from the 
de-excitation rate, which depends on neutral hydrogen density and temperature
that are computed self-consistently by following the relevant species in a 
non-equilibrium fashion.

Each photon launched from a cell is assigned a weight, calculated by dividing 
the total \lya luminosity of the cell by the number of simulation photons 
lauched from it. Such cell-dependent weights are accounted for in computing 
the \lya surface brightness profiles. The scatterings of the photon
with neutral hydrogen atoms on its way out and the corresponding changes
in position, direction, and frequency are tracked until it escapes the
grid boundary. We record the initial position of each photon, the position
of the last scattering, the direction and frequency after the last scattering,
and the fractional contribution of cooling radiation to its total luminosity. This information is used
to compute a mean surface brightness profile for each LAE, averaged over
all directions (see \S~2.2). 

At each scattering, we also compute the escape probability towards a fixed 
direction and collect the escaping \lya 
photons onto an integral-field-unit-like 3-dimensional array with pixel size 
the same as the cell size, which allows us to construct a \lya image of each 
LAE as viewed along the chosen direction \citep{Zheng:2002aa}.

Finally, we account for the effects of the intergalactic medium (IGM) outside 
of the box and the interstellar medium (ISM) in star-forming regions on 
\lya emission, following the approximate methods in \citet{Cen:2013aa}.

In brief, for each photon escaping the box at frequency $\nu$, we calculate 
the scattering optical depth $\tau_{\nu}$ from the edge of the box to an 
observer at $z=0$ using the redshift-dependent IGM hydrogen density and apply 
a factor of $e^{-\tau_\nu}$ correction for the IGM 
absorption. While such a correction neglects the differences in the IGM 
along different directions, it serves our purpose of introducing an overall 
average effect of the IGM.
We also apply an effective ISM dust attenuation to the intrinsic 
\lya emission by multiplying the luminosity represented by each simulation 
photon by a simple $e^{-\tau}$ factor, with $\tau = 
0.2[{\rm SFR}/(M_{\odot} $yr$^{-1})]^{0.6}$. This is loosely motivated by 
the observational trends of higher dust attenuation in galaxies with higher SFR
\citep{Brinchmann:aa, Zahid:2012aa}, and the power-law index follows the slope 
of the metal column density dependence on SFR in the simulation. 
The dust extinction can be thought as 
applied to the ionizing photons around the \ion{H}{2} regions, lowering the 
luminosity of \lya emission coverted from ionizing photons through 
recombination. The factor is also intended to absorb uncertainties in the
galaxy formation simulation (e.g., in the predicted SFR).
Our methods of applying IGM and ISM absorption 
are the same as adopted in \citet{Cen:2013aa}, in which the LAB luminosity
function and luminosity-size relation have been successfully reproduced, 
suggesting that the approximate treatments work well in capturing the major 
IGM and ISM effects and in absorbing model uncertainties.
In our current work of LAHs, we do not adjust any parameters and simply use
the direct outputs of the radiative transfer model to compare to observations.

We show in Figure~\ref{fig:nhimage} the column density distribution of neutral 
hydrogen gas around each of our model LAEs, viewed from a fixed direction.
The black contour curves are drawn at $10^{20.3}$cm$^{-2}$, delineating regions
corresponding to damped \lya systems (DLAs). DLAs represent regions 
extremely opaque to \lya photons, while \lya photons can be significantly 
scattered in regions of much lower column densities (e.g., 
above $10^{15}$cm$^{-2}$). The images reveal extended, 
filamentary structures of neutral hydrogen connecting regions of high column 
densities. Scatterings of \lya photons off hydrogen atoms in these structures
leave signatures in the resulting \lya surface brightness distributions.

Figure~\ref{fig:image} shows the corresponding \lya images of the nine LAEs 
in our analysis. The isophotal contours in each image 
correspond to $10^{-17} \SBunit$, about the surface brightness threshold for 
detecting individual $z\sim 3.1$ LAEs \citep{Ouchi:2007aa}. These images 
reveal a rich degree of structures and a variety of morphologies at fainter 
surface brightness levels, which allows the LAHs to be revealed by the stacking
analysis. The surface brightness distribution depends on the viewing angle
\citep[e.g.][]{Zheng:2010aa}. The stacked image (as in \citealt{Momose:2014aa})
for a sample of LAEs comes from averaging many images from galaxies of random 
orientations. For the relatively small number of galaxies modeled here, we can 
form the mean surface brightness distribution by viewing each galaxy from 
many different observing directions.

\subsection{Computing the Mean \lya Surface Brightness Profiles of Model LAEs}

\label{sec:SBmethod}

\begin{figure*}
\epsscale{1.17}
\plottwo{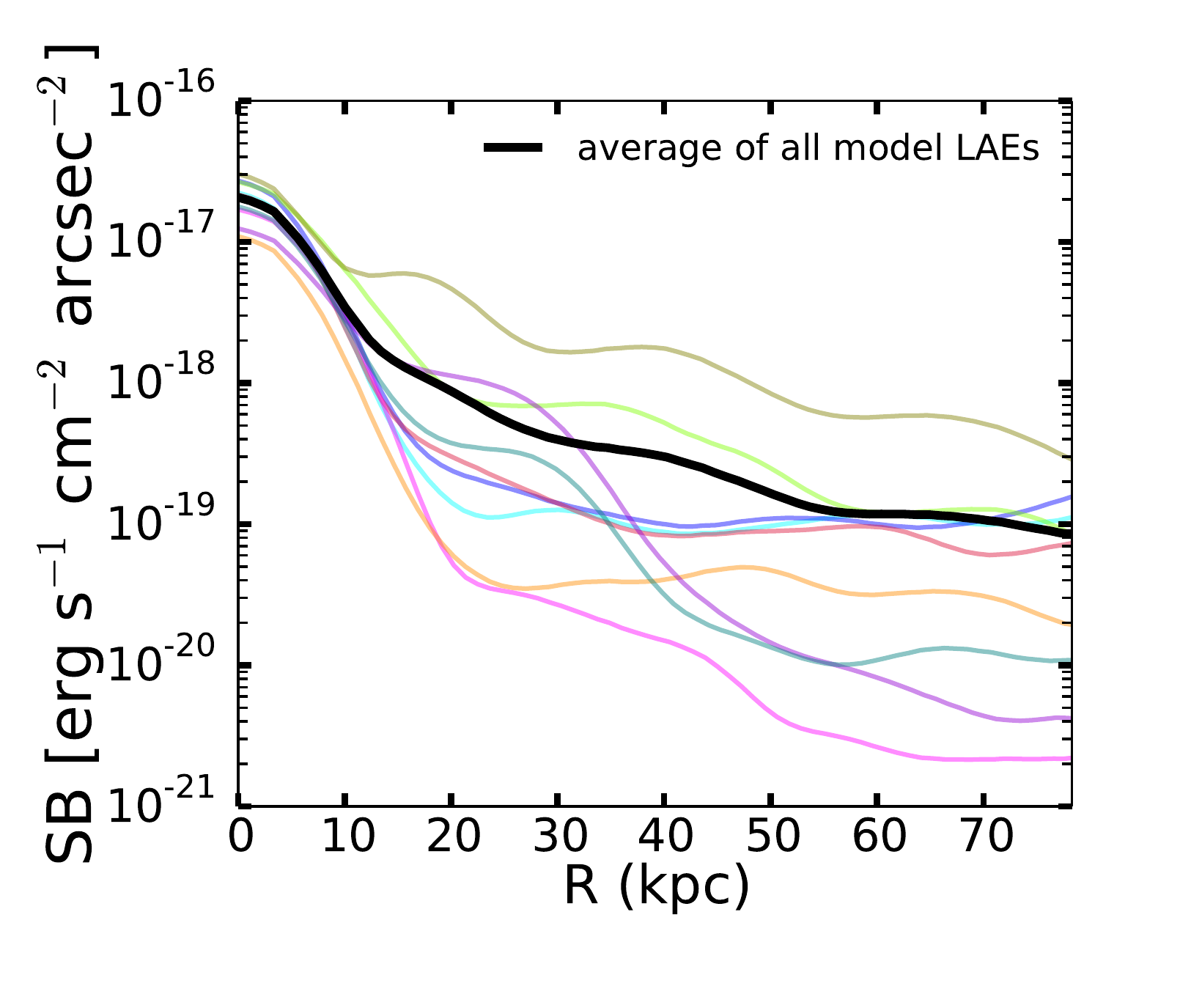}{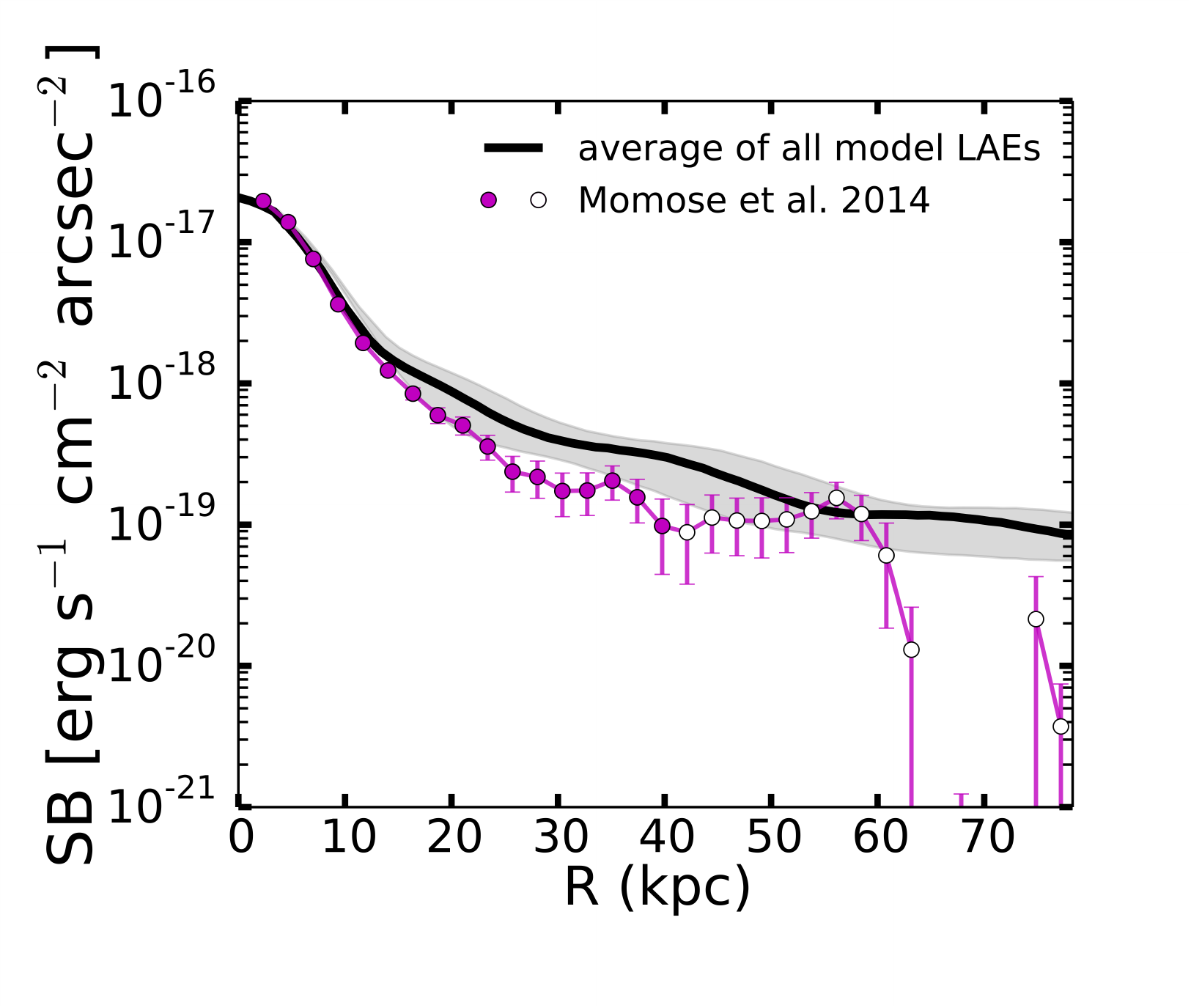}
\caption{
\label{fig:allprof}
\lya surface brightness profiles for the model LAEs in our analysis. 
{\it Left:} 
\lya surface brightness profiles for individual model LAEs,
with the average over all nine LAEs shown as the black curve. While the nine
LAEs have similar surface brightness levels at small radii, they display
large variations at large radii, reflecting the differences in gas 
distributions in our model halos. 
{\it Right:}
Comparison of the average \lya profile with observational results
in \citet{Momose:2014aa}. 
Beyond $\sim 40~\kpc$ systematic effects in the image stacking analysis
become important, indicated by the open circles. 
The shaded region gives an idea of the spread in the model profile,
obtained by excluding the LAE with the faintest or the brightest extended 
profiles from the average.
}
\end{figure*}

First, let us consider an observer located along a direction 
${\mathbf k}$
from one of our model galaxies. The average surface brightness ${\rm SB}(R,{\mathbf k})$ at a 
projected radius $R$ (physical) to the galaxy center as seen by the observer
can be computed as (assuming a spatially flat universe)
\begin{equation}
\label{eqn:SB0}
{\rm SB}(R,{\mathbf k})=\frac{\Delta L (1+z)^{-2}}{\Delta\Omega D_c^2 [A(1+z)^2/D_c^2]},
\end{equation}
where $\Delta\Omega$ is a small solid angle centered around ${\mathbf k}$, 
$A\equiv 2\pi R \Delta R$ is the area of a small annulus around $R$, and 
$\Delta L$ 
is the total luminosity of escaped \lya photons falling into 
$\Delta \Omega$ and with 
projected last scattering position within $A$. The quantity $\Delta L 
(1+z)^{-2} / (\Delta\Omega  D_c^2)$ is the corresponding flux 
(erg s$^{-1}$ cm$^{-2}$) the observer at a comoving distance $D_c$ receives,
with the two factors of $1+z$ from energy redshift and time dilation. The
quantity $A(1+z)^2/D_c^2$ is the solid angle extended by the annulus, seen by 
the observer, with the $(1+z)^2$ factor converting physical area to comoving.

The surface brightness profile at projected radius $R$ averaged over
all observing directions is then
\begin{equation}
\label{eqn:SB}
\langle {\rm SB}(R) \rangle = \frac{1}{4\pi} \int {\rm SB}(R,{\mathbf k}) d\Omega.
\end{equation}
Given Equation~(\ref{eqn:SB0}), for the annulus at a given $R$ and $\Delta R$, 
we only need to obtain the average of $\Delta L/\Delta\Omega$ over all 
observing directions for
computing the integral in Equation~(\ref{eqn:SB}). Denoting
the total \lya luminosity from this annulus
as $L_A$, we then have $L_A=\int \Delta L/\Delta\Omega\, d\Omega$. In the
limit of an infinite number of observing directions, we obtain
\begin{equation}
\langle {\rm SB}(R) \rangle = \frac{L_A}{4\pi A (1+z)^4}.
\end{equation}
This equation is the basis of computing the mean \lya surface brightness 
profile for each model LAE, along with the information we record for the escaping
\lya photons. For the mean \lya surface brightness at projected radius $R$, 
instead of producing images viewed from many directions, we only need to 
obtain the sum of the total \lya luminosity $L_A$ for photons whose 
projected radii fall into the annulus around $R$ ($R\pm \Delta R/2$ with 
an area $A$). The projected radius $R_\gamma$ of a photon is computed from 
its escaping direction ${\mathbf k}_\gamma$ and its position of last 
scattering ${\mathbf r}_{\rm ls}$ (with respect to the galaxy center) as
$R_\gamma = \sqrt{ r_{\rm ls}^2 -({\mathbf k}_\gamma \cdot {\mathbf r}_{\rm ls})^2 }$.
We have verified that the method gives the same results as that from averaging 
images over many observational directions.

Our simulation has a much higher resolution (cell size of $\sim$0.04\arcsec) 
than the observation in \citet{Momose:2014aa}. To mimic the smoothing effect 
in the images of \citet{Momose:2014aa}, we obtain the final surface brightness 
profile by convolving the resulting \lya image with a 2D Gaussian kernel 
with a full width at half maximum (FWHM) of 1.32\arcsec, corresponding to 
10.3~\kpc (physical) at $z=3.1$.

\section{Results}

We first present the results on the mean \lya surface brightness profile in 
our model. We then decompose the mean profile in various ways to study its origin. Finally,
we compare our UV profile to observations in order to further constrain the relative 
contributions of cooling and star-forming emission.

\subsection{The Mean \lya Surface Brightness Profile}

The left panel of Figure~\ref{fig:allprof} shows the mean surface 
brightness profiles from the smoothed images of the nine model LAEs in 
our analysis. The overall mean
of the nine profiles is plotted in black. For each LAE, the central profile
(e.g., $R\lesssim 15~\kpc$) is largely determined by the smoothing kernel 
(point spread function; PSF). The
peak surface brightness at the center shows a small variation among the
nine individual profiles, around (1--3)$\times  10^{-17}\SBunit$.

At large projected radii, each LAE shows an extended profile, which can be
identified as the diffuse LAH. The profile at large radii is much flatter than
the central part. The surface brightness level of this extended component 
displays a substantial variation among the nine individual LAEs, as large as 
two orders of magnitude.

In the right panel of Figure~\ref{fig:allprof}, we compare the overall mean 
profile of the nine model LAEs with the one derived by \citet{Momose:2014aa} 
from stacking the \lya images of 316 $z\sim 3.1$ LAEs. According to 
\citet{Momose:2014aa}, the data at $R \lesssim 40~\kpc$ are reliable, while
at larger radii systematic effects in the stacking analysis become significant 
compared to the signal (see the top-middle panel in their Figure 8). We mark 
such a transition by using filled circles at $R \lesssim 40~\kpc$ and open 
circles at $R \gtrsim 40~\kpc$ for the data points. The shaded region around 
the mean model profile quantifies the uncertainty. The upper (lower) 
boundary is derived by excluding the LAE with the lowest (highest) surface 
brightness and averaging over the other eight LAEs. This serves to only 
provide some rough idea on the variation level of the mean profile, given the
small number of model LAEs in our analysis.

On small scales ($R\lesssim 15~\kpc$), the model profile matches the 
observed profile extremely well, which is striking. At first glance one may
attribute this to coincidence, since we do not intend to fit the observed
profile and we do not have any free parameters to adjust in our model. 
We directly use the \lya emissivity and gas distribution in the simulation.
The only two changes we apply in the model besides the radiative transfer 
calculation are ``effective'' dust and IGM absorption. The ``effective''
\lya extinction optical depth is the same as the one adopted in \citet{Cen:2013aa}
for studying LABs, which suppresses the initial intrinsic \lya emission. 
It aims at accounting for any uncertainties in the galaxy formation 
simulation. We tie it to the star formation rate and the relation is fixed
by considering halos above $10^{12}\Msun$ in \citet{Cen:2013aa}. We apply a
mean absorption (scattering) for \lya photons escaping the grid from the IGM 
outside of the simulation box. 

In \citet{Cen:2013aa}, the observed \lya luminosity-size relation of LABs 
and the \lya luminosity function of LABs are reproduced by our radiative 
transfer modeling. So it may not be too surprising that the similar model 
also provides a good match to the \lya emission in lower mass halos 
($10^{11.5}\Msun$). Since the surface brightness profile at the central 
part is largely determined by the PSF, the agreement with the observation
means that the central \lya luminosity in our model happens to be similar 
to the average luminosity of the 316 LAEs in the stacking analysis in
\citet{Momose:2014aa}, which by all means is an encouraging sign. 

At larger radii ($R\gtrsim 15~\kpc$), the model curve is slightly higher 
(at a factor of two level) than the observation. Given the small number of
LAEs and the lack of adjustment in the model, the agreement to the 
observation still appears remarkable, in particular if the uncertainties in the
data points and in the model curve are taken into account (keeping in mind that
the data may suffer from significant systematic bias at $R \gtrsim 40~\kpc$).
Both the shape and extent of the LAH are reasonably reproduced.

As a whole, our model mean \lya surface brightness profile, effectively 
computed from
stacking \lya images of nine LAEs viewed along many different directions, shows
good agreement with stacking analysis from observed $z \sim 3.1$ LAEs, 
from the 
central part to the diffuse LAH extended to $R\sim 60~\kpc$. However, the 
systematic effect in the data analysis makes the comparison beyond 
$\sim 40~\kpc$ less reliable, and the apparent disagreement at 
$R \gtrsim 60~\kpc$ is not significant.
We proceed to investigate the contributions from various components to the surface brightness
profile to gain more insights.
 
\begin{figure*}
\epsscale{1.17}
\plottwo{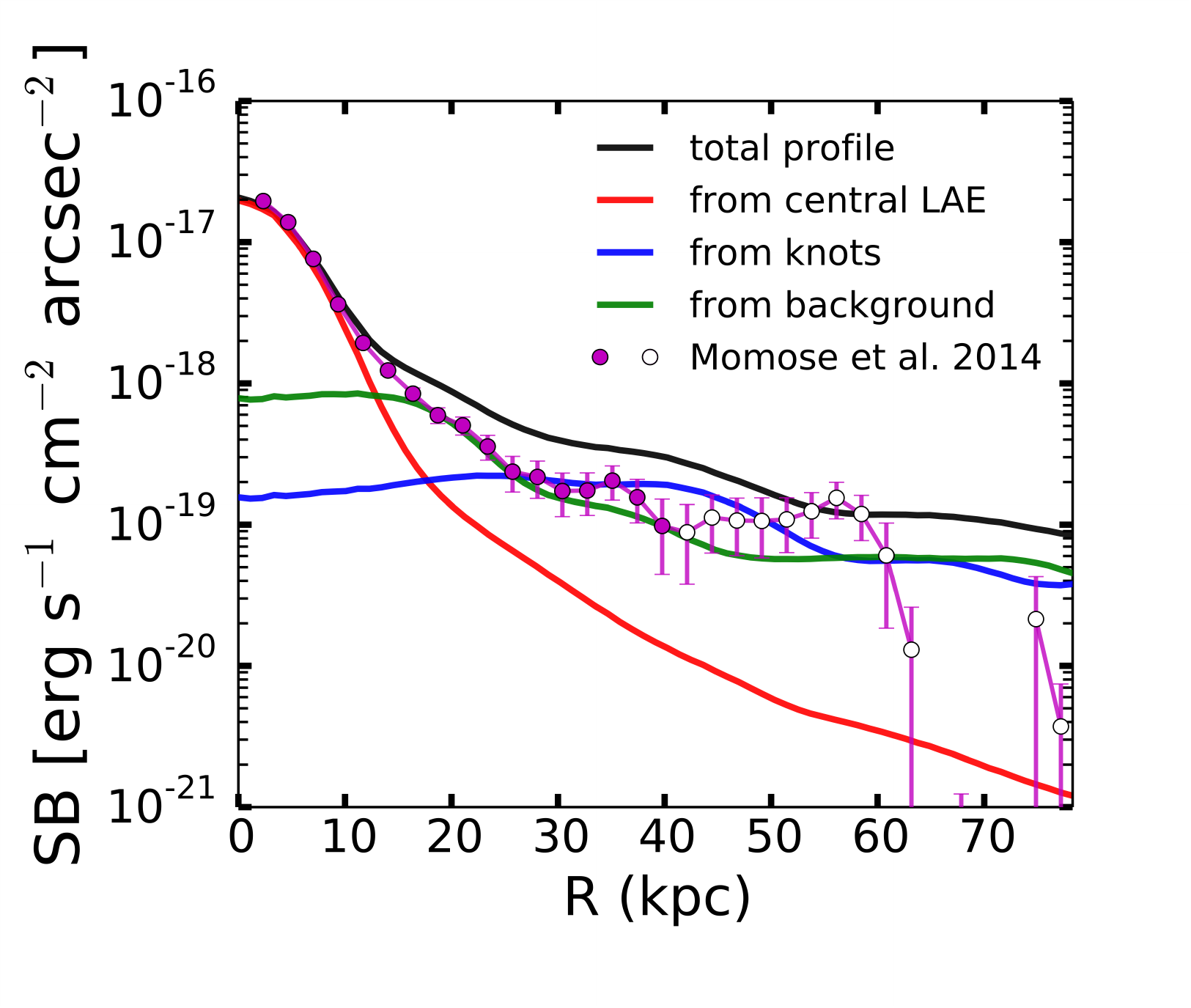}{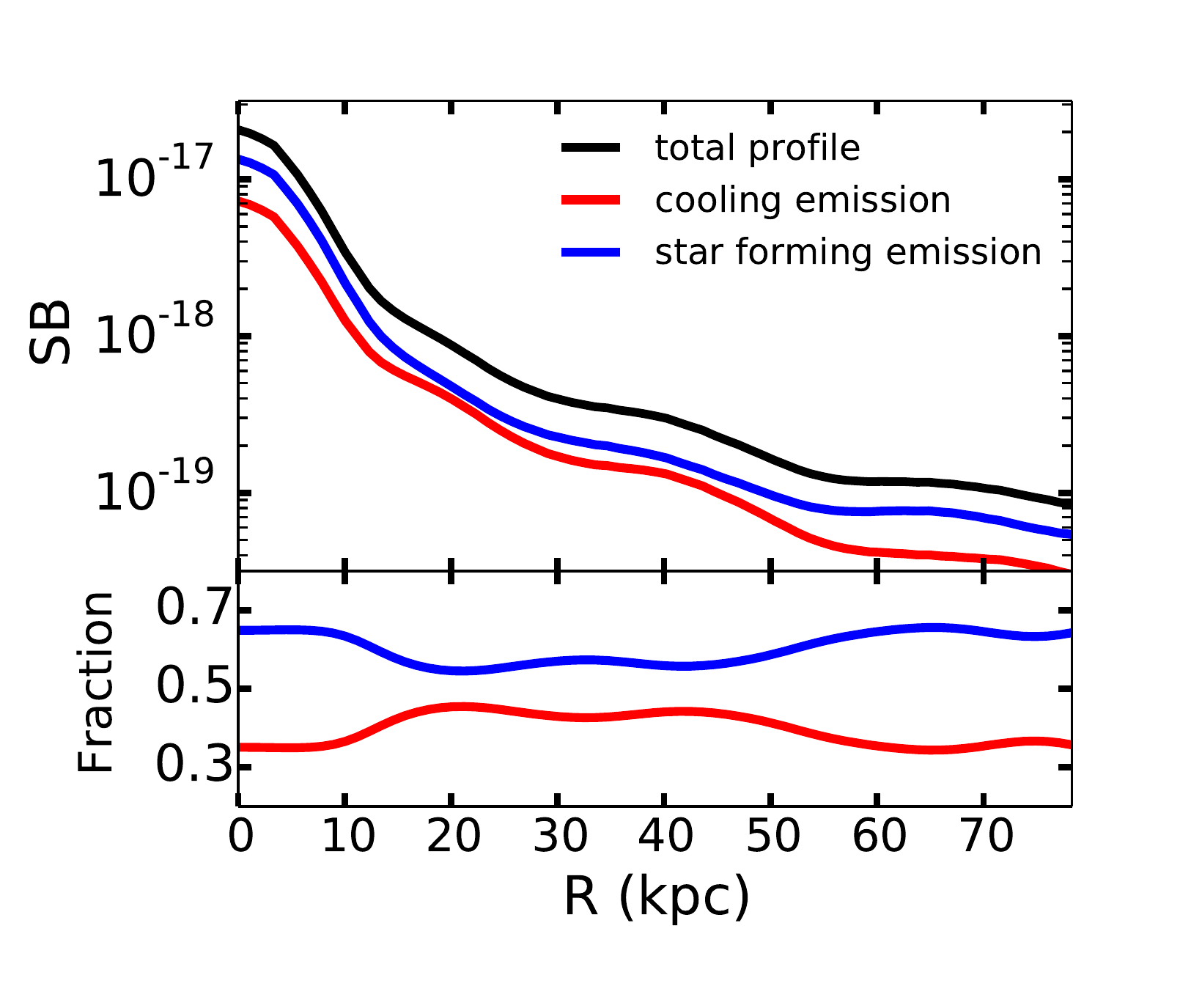}
\caption{
\label{fig:decomp}
Decomposition of the total \lya surface brightness profile into different 
components of Ly$\alpha$ emission.
{\it Left:}
The decomposition of the total profile into contributions from the central LAE (red curve), 
other high emission knots (blue curve), and background regions (green curve). Observational data from 
\citet{Momose:2014aa} is shown in purple.
Systematic bias is important for $R \gtrsim 40~\kpc$, indicated by the open 
circles.
Note that at radii larger than $\sim 10~\kpc$, the profile from 
the central LAE is unable to account for observations, with the knot and background profiles playing 
dominant roles.
{\it Right:}
The decomposition of the total profile into contributions from star formation and cooling emission. The top panel 
shows profiles for each emission type, given in surface brightness units of erg s$^{-1}$ cm$^{-2}$ arcsec $^{-2}$. 
Star forming emission is shown in blue, and cooling emission is shown in red. The bottom panel shows 
the fractional contribution that each emission type makes to the total profile.} 
\end{figure*}

\subsection{Decomposing the \lya Surface Brightness Profile}

\begin{figure*}
\epsscale{1.17}
\plottwo{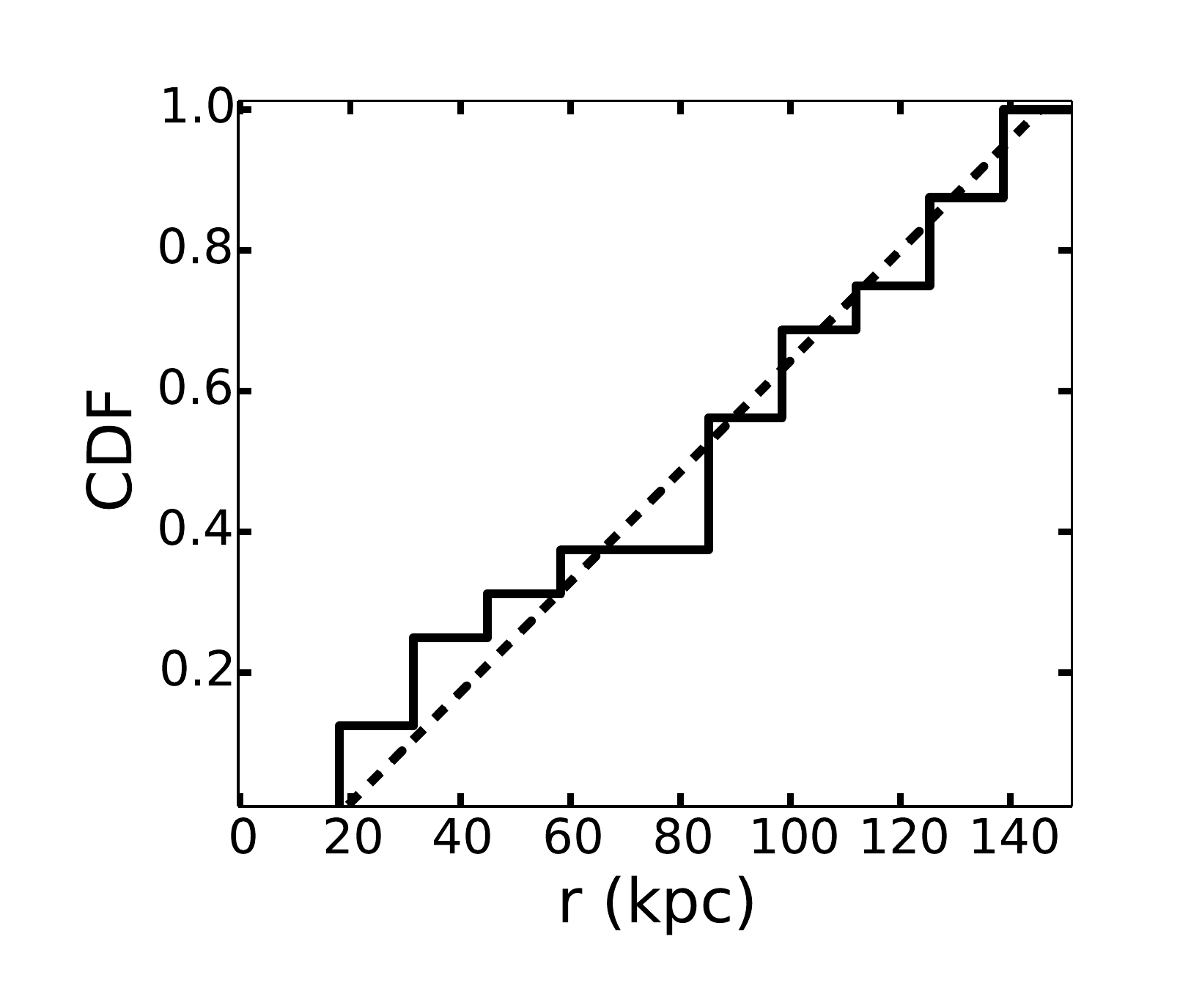}{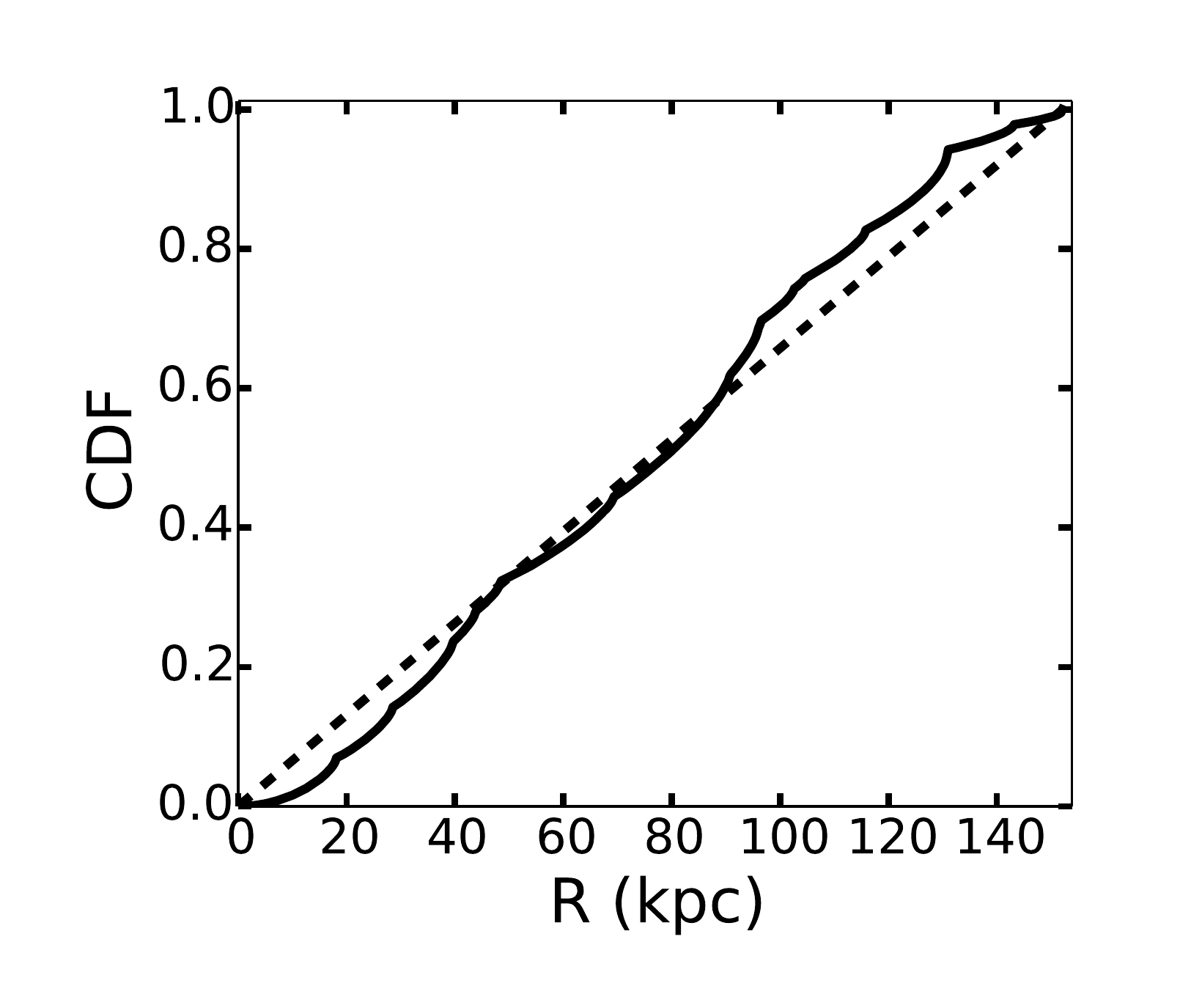}
\caption[]{
\label{fig:knots_cdf}
Spatial distribution of the identified high emission knots in our simulation. 
{\it Left:}
The distribution of 3D distances of the knots to the box center. The solid line 
shows a cumulative distribution function 
(CDF) of the 3D distances, with the dashed curve 
corresponding to a number density profile $ n(r) = 0$ for $r < 20~\kpc$ and 
$n(r) \propto r^{-2}$ for $ r \geq 20~\kpc$. 
{\it Right:}
The distribution of 2D projected distances of the knots to the box center. The 
solid line shows the CDF of the 2D distances 
and the dashed curve is for a number density distribution $\propto R^{-1}$. 
}
\end{figure*}

\begin{figure*}
\epsscale{1.17}
\plotone{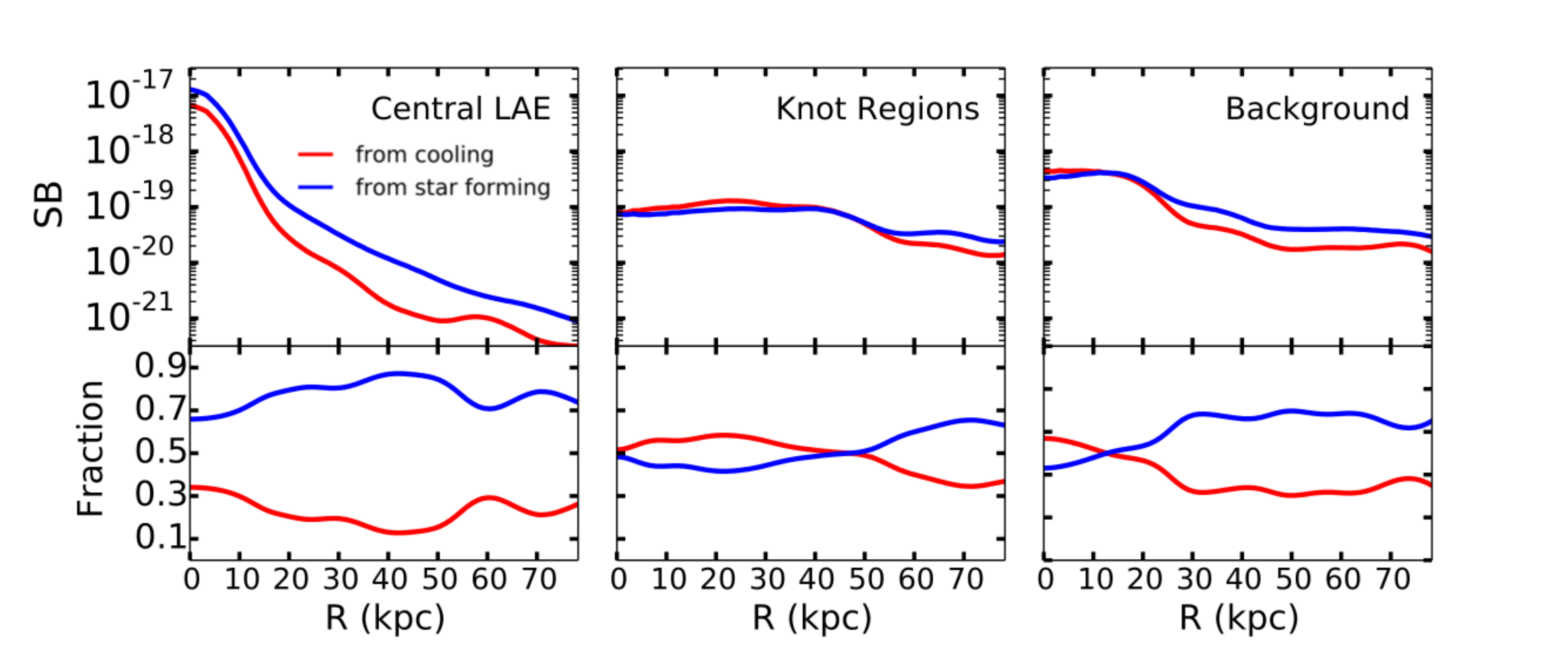}
\caption{
\label{fig:sfcooling}
A further decomposition of the averaged surface brightness profile into 
contributions from star formation (blue curves) and cooling radiation 
(red curves). From left to right, the panels show profiles for photons 
produced within 10~\kpc\ of the central LAEs, within the knot regions, 
and from the background areas. The top panels show surface brightness in 
units of $\SBunit$, while the bottom panels show the relative contributions.
}
\end{figure*}

We record the initial position of each photon, which makes it possible to 
separate the contributions to the surface brightness profile from photons
originating at different places in our simulation. 

In each halo, there is a central LAE with strong \lya emission. We attribute 
\lya photons launched within 10~\kpc\ of the halo center to the central LAE. There 
are also a few regions in the halo with high \lya emissivity, which are small
star-forming galaxies around the central LAE. In the \lya images shown in 
Figure~\ref{fig:image}, such regions appear as relatively isolated peaks 
with surface brightness above $\sim 10^{-19} \SBunit$. The majority of
them are below the detection threshold for typical LAE surveys, as is the
case in \citet{Momose:2014aa}. In three of the nine LAEs, a few of the high 
emissivity regions can reach the detection threshold, and would show up
as isolated LAEs around the central LAEs. We refer to the high emissivity 
regions as ``knots'' and associate to each knot \lya photons launched within
10~\kpc\ of its center. Photons that belong to 
neither the central LAE nor the knots are identified as being emitted from 
the background of the simulation box. Most of them come from small clumps of gas
that possess low rates of star formation. 
Clearly the distinction between the knots and the background depends on our choice, which
can be arbitrary. However, our separation here serves the purpose of obtaining
a rough idea on how \lya photons from different physical regions contribute 
to the surface brightness profile.

The left panel of Figure~\ref{fig:decomp} shows the decomposition of
the mean surface brightness profile (black) into contributions from 
\lya photons originating in the central LAE (red), the knots regions (blue),
and the background regions (green).

After the radiative transfer, \lya photons originating from the central LAE 
appear to peak around the central region, following the PSF. The entire
amplitude of the overall surface brightness profile at $R<10~\kpc$ comes from 
this component. The scatterings of photons with neutral hydrogen atoms in the 
circumgalactic and intergalactic media lead to an extended profile beyond 
$\sim 15~\kpc$. The profile drops toward large radii, roughly following 
$R^{-3.3}$, which is too steep to account for the LAHs seen in both the model 
and observations.

\lya photons from the knots and background regions make comparable 
contributions (within a factor of about two) to the overall surface 
brightness profile at scales above $\sim 20~\kpc$. They have similar
profiles, which in turn are similar to that of the LAH and are flatter than 
the extended profile from scattered photons from the central LAE. Together, 
they dominate the profile at $R\gtrsim 20~\kpc$.

The above decomposition leads to the interesting implication that
the observed extended \lya emission of LAHs is largely caused by emission 
from regions of low star formation rates spatially distributed inside the 
host dark matter halos of the central LAEs, and that photons diffusing 
out from the central LAE as a result of the radiative transfer process 
play only a secondary role in producing the observed extended emission.

We also decompose the surface brightness profile into contributions from
\lya photons generated by star formation and cooling radiation, as shown in 
the right panel of Figure~\ref{fig:decomp}. The profiles from the two 
contributions are similar, but the photons from cooling radiation always
sub-dominate, making about 30-40\% of the total \lya light in the extended 
LAHs. 

Because the identified high-emission knots contribute significantly to the 
surface brightness profile at large radii, it is important to examine their 
spatial distribution.  The left panel of Figure~\ref{fig:knots_cdf} shows
the cumulative distribution function (CDF) of the 3-dimensional (3D) distance 
$r$ from the knots to the center for all the nine LAEs. Since we define the 
radius of 
each knot region and the central LAE to be 10~\kpc, any high-emission areas 
within 20~\kpc of the center will not be identified as independent knots in 
our analysis. Therefore, the CDF curve starts at $r=20~\kpc$. The dashed curve 
corresponds to a number density distribution $n(r)\propto r^{-2}$ (for 
$ r \geq 20~\kpc$ and 0 for $r < 20~\kpc$). The plot shows that the 
high-emission knots closely follow a singular isothermal distribution up to 
at least $\sim 3\Rvir$.

As we study the surface brightness profile, it would be more illustrative to
examine the projected distribution of the knots. For this purpose, we chose a
large number of isotropically distributed viewing directions. For
each viewing direction and each LAE, we record the projected radius to the 
center for each knot. The right panel of Figure~\ref{fig:knots_cdf} shows the 
CDF for the 2D projected radii $R$. The dashed curve is the CDF for a
surface number density that follows $R^{-1}$, which appears to be a reasonable 
description of the distribution of knots. This is consistent with the CDF of 
the 3D radii. The distribution of knots explains the shallow slope in the
mean surface brightness profile seen in the model or observed LAHs (e.g., 
Figure~\ref{fig:allprof} and Figure~\ref{fig:decomp}), which has a slope 
around $-1$.

Figure~\ref{fig:sfcooling} shows the decomposition of our surface
brightness profiles into contributions from star formation and cooling 
radiation for the central LAE, the identified knot regions, and the background. 
For \lya photons produced in the central LAE, cooling radiation makes up about 
1/3 of the observed \lya emission near the center, and its contribution drops 
to 15\% at $R\sim 40~\kpc$. For \lya photons produced in the knot regions, 
cooling radiation and star formation contributions are comparable, and
for those produced in background, the star formation contribution dominates. 
The latter two components depend on how we define knot regions. If we choose
a lower threshold to define knots, star formation would become the dominant
mechanism in producing \lya photons in knot regions.

\subsection{Possible Constraints from the UV Surface Brightness Profile}

\begin{figure*}
\epsscale{0.93}
\plotone{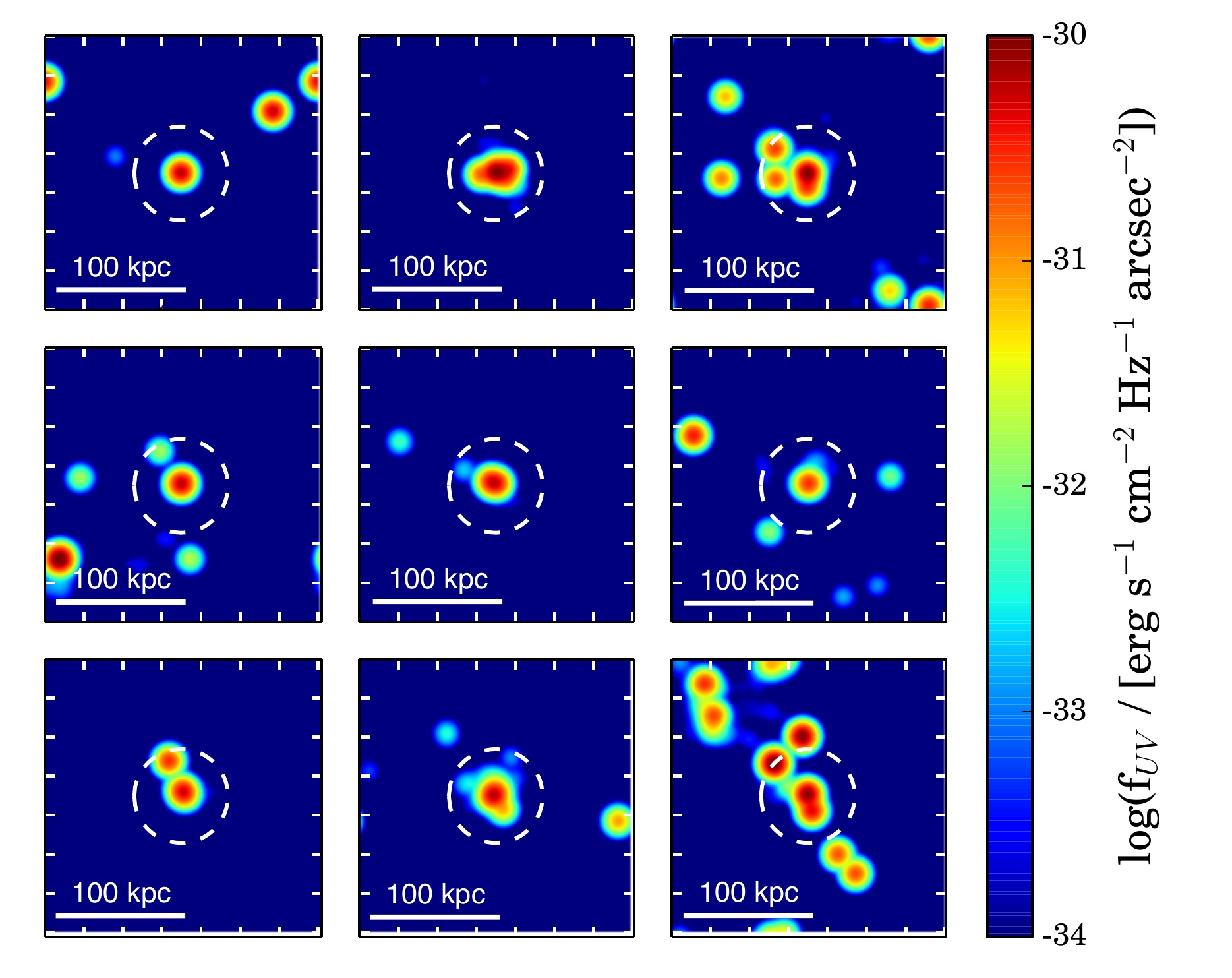}
\caption[]{
\label{fig:UVimage}
UV surface brightness images for all nine model LAEs in our original analysis. 
Each image is 224~\kpc\ (physical) on a side and has been smoothed by a 2D 
Gaussian kernel with a FWHM of 1.32\arcsec\ to match observations. The dashed
circle in each panel has a radius of 40~\kpc, roughly corresponding to the 
radius that systematic effects become important in the stacking analysis in 
\citet{Momose:2014aa}. The observed profile can potentially be used to put 
constraints on the clustered UV sources around the central galaxies, as 
discussed in the text. 
}
\end{figure*}

The \lya surface brightness profile from our model shows an encouraging 
agreement with the data. Besides \lya, the stacking analysis is also performed
for UV images \citep[e.g.,][]{Momose:2014aa}. UV photons are produced from star
formation. Unlike \lya photons, they do not interact with neutral hydrogen
through resonant scattering. Instead, they escape directly from their point of 
creation (modulated by dust extinction), which allows them to serve as a tool 
to map out regions of star formation. The UV profile of LAEs can therefore
provide complementary information about LAHs and can be used to further
constrain the origin of LAHs.

We convert the SFR to UV luminosity (at rest-frame 1500\AA) using the 
prescription $L_{UV} = 8 \times 10^{27}[{\rm SFR}/(\Msun {\rm yr}^{-1})] 
{\rm erg\, s^{-1}Hz^{-1}}$ \citep{Madau98}. It assumes a 
Salpeter initial mass function (IMF) of stars and solar metallicity. 

While 
dust extinction should also be included to obtain the observed UV luminosity, 
it is degenerated with the above assumption about the stellar population
and to a less degree with the star formation history. For 
example, a sub-solar metallicity (or a different IMF, e.g., Chabrier
IMF) results in a higher SFR-to-UV conversion factor, and the combined effect
of the metallicity and IMF can lead to a factor of a few increase in the
conversion \citep[e.g.,][]{Leitherer99,Madau14}. The dust extinction works
in the direction to bring the conversion factor toward the above value.
Given such uncertainties in the model, we can simply adopt the above 
conversion to proceed, and rescale the model UV profile if necessary to 
fit the observed profile.

We create UV photons based on the SFR in each cell and 
assign a random escape direction for each photon. Figure~\ref{fig:UVimage} 
displays the UV images of the nine model galaxies observed along the 
same direction as in Figure~\ref{fig:image}. 
To study the average UV light distribution seen in the stacking analysis, we 
follow the method described in \S~\ref{sec:SBmethod} to produce the average 
UV surface brightness profile for each model LAE.

\begin{figure*}
\epsscale{1.17}
\plottwo{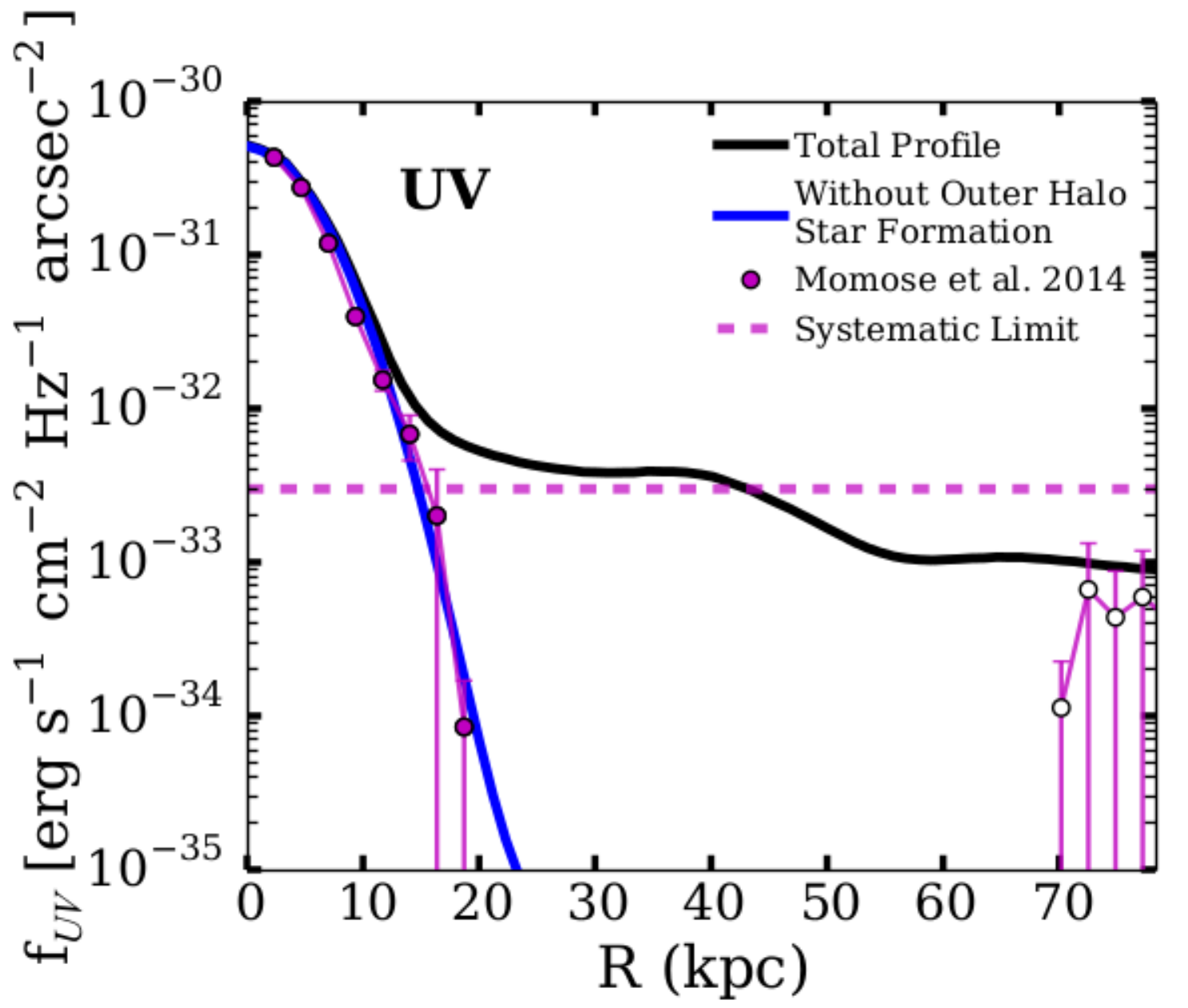}{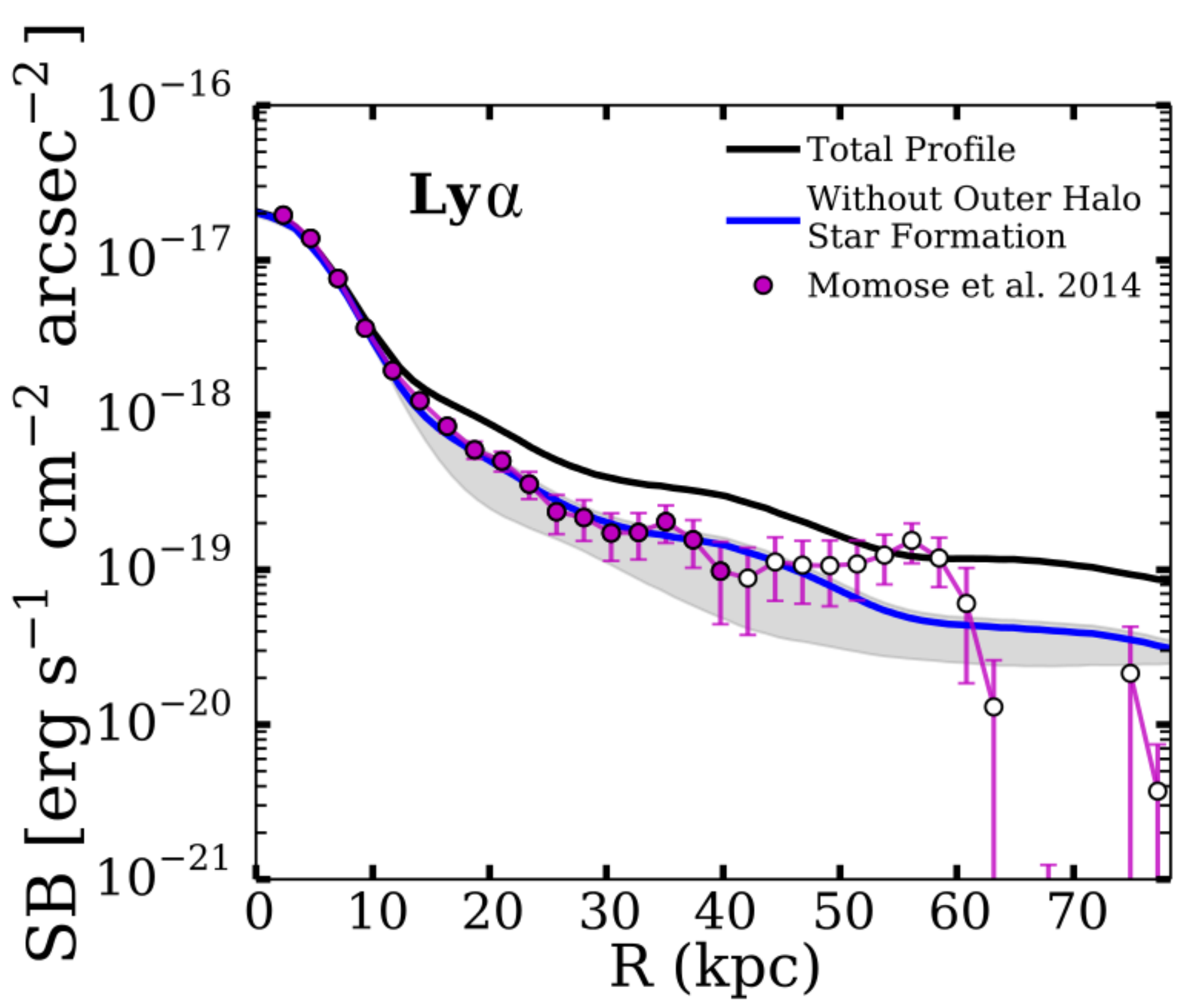}
\caption{
\label{fig:UVprof}
Average UV and \lya surface brightness profiles and the effect of star 
formation in the outer halo. In each panel, observational data are taken from \cite{Momose:2014aa}
with open circles representing the region where the data starts to be 
limited by systematics.
{\it Left:} Comparison of our UV profiles with observational data. The black curve
is the average profile from our initial model. The blue curve shows the effect 
of removing star formation in the outer halo, keeping only 
the star forming emission from the central LAE. The dashed
line is drawn at $3.0\times10^{-33}\UVunit$, which is the level of systematic 
effects in the image stacking analysis (see the text). This line represents 
an upper limit for any extended UV emission that actually occurs. 
{\it Right:} The corresponding changes in the model \lya profile. The black curve shows 
our total \lya profile, and the blue curve shows the effect of removing star formation
emission from the outer halo. The gray shaded region in the right panel gives an idea of the spread of the
profile, obtained by excluding the LAE with the faintest or the brightest 
extended profiles from the average.
}
\end{figure*}

The black curve in the left panel of Figure~\ref{fig:UVprof} shows the UV 
profile predicted by our model. At small radii ($R\lesssim 15~\kpc$), the 
model curve is almost right on top of the observed profile. 

Such a coincidence 
indicates that our SFR-to-UV conversion factor is about right in reflecting
a combined effect of stellar population, metallicity, extinction, and model 
SFR, even though there are uncertainties in each component and the overall
model is approximate. As such, we make no adjustments to our initial conversion factor.

The model reproduces
the central UV luminosity, and the shape of the profile simply follows that 
of the PSF. 

At large radii ($R\gtrsim 15~\kpc$), our model shows an extended UV halo
(left panel of Figure~\ref{fig:UVprof}).  This is not unexpected, given that 
emission produced from star formation in the outer halo makes a substantial 
contribution to the extended \lya profile (Figure~\ref{fig:sfcooling}). 
The model UV profile is in apparent tension with observations 
\citep{Momose:2014aa}, where little  evidence is shown for such extended UV 
halos (see the data points in the left panel of Figure~\ref{fig:UVprof}). 

We note that there are some residuals of sky subtraction found in the 
composite UV images of \citet{Momose:2014aa}. To quantify the significance 
of the apparent tension between the model and observation, we evaluate the 
sky subtraction systematics in the average UV surface brightness profile.
We find that there is a signal of sky over-subtraction at the surface 
brightness level of $3.0\times10^{-33}\UVunit$, corresponding to a maximum
negative value of the UV surface brightness estimate of the LAH (Momose et al.
2015, in preparation). Therefore, we conclude that the average UV profile
below the above level is subject to the influence of the systematics. 
Note that the effect of sky over-subtraction systematics is canceled out in 
the \lya surface brightness profile of the LAH, because the \lya profile is 
obtained by taking the difference of the composite broadband and 
narrowband images, which have the same level of sky over-subtraction.

The dashed line in the left panel of Figure~\ref{fig:UVprof} shows the 
sky-subtraction systematic effect at the surface brightness level of 
$3.0\times10^{-33}\UVunit$, which can be regarded as an upper bound for 
extended UV profile. Our model UV profile (black curve) appears to be at a 
similar level. Therefore, by accounting for the systematics in the data,
no significant tension is found between the model and the observation.
Given the sky-subtraction systematics, the UV profile obtained in 
\citet{Momose:2014aa} (data points in the left panel of our 
Figure~\ref{fig:UVprof}) represents a lower bound, which essentially follows
the PSF and is determined by the central star formation. To match this case,
we construct a modified model by removing star formation in the outer halo 
regions in our model. It is
not surprising to see that the modified model profile produced in this way
(blue curve in the left panel of Figure~\ref{fig:UVprof}) follows the PSF, 
as with the data points. 

The default model profile (black curve) and the modified one (blue 
curve) should be able to bracket all possible cases of the extended UV
profile. Improved measurements of the UV profile with well-controlled 
systematics would provide important information on the amount of star
formation in the outer halo. This in turn would improve our understanding
of the origin of the extended \lya halo, along with its partition into contributions
from star formation and cooling radiation. With the current situation,
the above two boundary cases allow us to infer the range of the relative 
contributions of star formation and cooling radiation to the \lya surface 
brightness profile. 

The black curve in the right panel of Figure~\ref{fig:UVprof} corresponds to
the \lya profile from our default model. We find that the total \lya luminosity
within the projected radius $R<R_{\rm vir}=56~\kpc$ can be broken down into 
the following contributions: 33\% from star forming photons produced in the 
central galaxy, 15\% from cooling radiation emitted from the central galaxy 
($r<10~\kpc$), 28\% from star forming photons produced in the outer halo, and 
25\% from cooling radiation in the outer halo. In the extended part of the 
profile (defined as emission observed at $15~\kpc \lesssim R \lesssim 
56~\kpc$), the fraction of photons from cooling radiation is about 42\%.

The \lya profile after removing star formation from the outer halo is
shown as the blue curve in the right panel of Figure~\ref{fig:UVprof}.
The effect is not drastic, and the \lya profile for this case drops to a 
level in even better agreement with observations. For this modified profile, 
the average component contributions to the total \lya luminosity within 
projected radius $R<R_{\rm vir}=56~\kpc$ are 45\% from star forming photons 
produced in the 
central galaxy, 20\% from cooling radiation within and around the central 
galaxy ($r<10~\kpc$), and 35\% from cooling radiation in the outer halo. 
On average, cooling radiation can now contribute to about half of the 
total \lya luminosity. If we focus on the extended part of the profile 
($15~\kpc \lesssim R \lesssim 56~\kpc$), the fraction of the cooing radiation 
contribution is about 75\%. With this prescription, the extended \lya halo is 
dominated by cooling radiation. 

Although the best currently available data is limited by systematics, 
the UV profile in combination with the \lya profile can 
help to constrain the nature of LAHs. Our investigation implies that
cooling radiation in the outer halo may play a significant role in forming
extended \lya halos (e.g., contributing more than half of the emission), 
a prediction that can be tested with tighter observational constraints on 
the UV profile. The caveats and more discussions are presented in the
next section.

\section{Summary and Discussion}

We perform \lya radiative transfer modeling of $z=3.1$ LAEs with a high 
resolution hydrodynamic cosmological galaxy formation simulation to study
the extended Ly$\alpha$-emitting halos recently discovered in observation from
image stacking analysis. We develop a method to compute the mean surface 
brightness profile from averaging over many different viewing directions. 
We consider nine model LAEs residing in halos of $10^{11.5}\Msun$ and find their
mean \lya surface brightness profile to be in remarkable agreement with the 
observed profile in \citet{Momose:2014aa}, at both the central and
extended parts.

To investigate the origin of the extended \lya emission, we decompose the 
profile into contributions from \lya photons produced in different regions, 
which include the central LAE in each simulation box, dense regions of 
high star formation activity spatially separated from the central LAE 
(dubbed as ``knots''), and faint background areas. The latter two outer halo
components are associated with satellite galaxies or tidally stripped 
materials in the halo. 

\lya photons originating near the halo center (from both star 
formation and cooling radiation) but scattered to large radii by the 
hydrogen atoms in the CGM do produce an extended \lya halo, as predicted by,
e.g., \citet{Laursen:2007}\footnote{Cooling radiation in the outer 
halo is also included in \citet{Laursen:2007}. However, there is no 
discussion on its contribution to the LAH for an obvious reason -- LAHs had not
yet been discovered in observations.}
and \citet{Zheng:2011aa}.
However, our radiative
transfer model with the high-resolution galaxy formation simulation 
shows that such a contribution 
alone is not able to explain the surface brightness level of the observed 
LAH (e.g., lower by a factor of 10 around $R=40~\kpc$). Instead, the extended 
LAH is dominated by emission from the knots and the background regions of the 
outer halo. The result implies that scattering of \lya photons from bright, 
central sources is less important in forming LAHs than previously 
thought.

Certainly the exact profile created by the scattered, centrally produced 
photons should depend on the density and velocity distribution of the 
circumgalactic gas. Some analytic models with clumpy CGM and decelerating 
outflows can produce scattered \lya halos at the observed surface brightness 
level \citep[e.g.,][]{Steidel:2011aa,Dijkstra:2012aa}. It is worth
 investigating the contribution from such scattered halos with more realistic 
CGM distributions from high-resolution galaxy formation simulations with 
various prescriptions of the star formation feedback 
\citep[e.g.,][]{Suresh:2015aa,Muratov:2015aa,Kimm:2015aa}. 
By performing a test with a lower-resolution grid used in the radiative 
transfer calculation, we also find that the scattered profile shows a weak 
dependence on the resolution, becoming slightly more extended
with higher resolution. However, the scattered profile still drops rapidly 
toward large radii, remaining unlikely to account for the observed one.

\lya emission from the outer halo has two components -- gravitational 
cooling radiation and emission from star formation. In the simulation, we
find that star formation slightly dominates, and cooling radiation makes a
substantial contribution. Our model predicts the existence of 
an extended UV halo at a brightness level of $\sim3\times10^{-33}\UVunit$, 
which is right at the limit of the sky-subtraction systematics in observational data
\citep{Momose:2014aa}. If actual UV halos (from improved data analysis) are 
significantly dimmer than this, we will need to investigate how to suppress 
the UV profile in the model (e.g., identifying the likely cause in the
simulation for the 
over-prediction of star formation in the outer halo or studying the dust 
attenuation effect in the outer halo). Although the best currently available 
UV profile measurement does not serve as a robust constraint as a result of 
 systematics, we find that the extended \lya profile becomes even better 
reproduced after removing star formation in the outer halo in order to 
suppress the UV profile.

Taken at face value, our investigation shows that our initial model can explain 
both the observed \lya and UV profiles of LAEs in \citet{Momose:2014aa}, from
small radii ($\lesssim 15~\kpc$) to large radii (up to $\sim 80~\kpc$). The 
agreements between the model and data are excellent. This is remarkable, 
especially given that we do not intend to fit the profiles by tuning 
parameters. Depending on the accuracy of our star-forming recipe and the 
loose constraints from the UV halo, we find that cooling radiation can 
contribute 40-55\% of the total \lya luminosity within
projected radius $R<R_{\rm vir}$, where $R_{\rm vir}=56~\kpc$ 
($\sim 7.2\arcsec$) is the virial radius of the LAE host halo. 
For the diffuse LAH, which is usually buried in sky noise for individual LAEs, 
the contribution from cooling radiation is more substantial, making up about 
42-75\% (within $15~\kpc \lesssim R \lesssim 56~\kpc$).

Gravitational cooling radiation from accretion of gas is a process expected 
to occur during galaxy formation, mainly in the form of \lya emission from 
collisional excitation and de-excitation of hydrogen atoms in gas around 
$2\times 10^4$ K \citep[e.g.,][]{Fardal01}. Many previous studies of cooling
radiation with analytic calculations and hydrodynamical simulations focus 
on investigating it as a possible mechanism to explain \lya blobs 
\citep[e.g.,][]{Haiman00,Fardal01,Furlanetto:2005aa,Yang06,Dijkstra:2009,
Goerdt10,Faucher10}, which are more luminous than
LAHs \citep{Steidel00}. As shown by \citet{Yang06} and \citet{Faucher10},
an accurate prediction of the cooling \lya emission relies on an accurate
treatment of the self-shielding effect for the ionizing photons, which affects
the ionization and thermal states of the accreted gas. In the simulation used
in this work, self-shielding correction is performed on-the-fly. There could be
small variations in the predicted cooling radiation with different 
self-shielding correction methods, but it is unlikely to remove the cooling 
radiation signal in our model, which is significant regardless of the 
accuracy of our star-forming recipe. 

One caveat to keep in mind is that our results in this paper are based
on radiative transfer modeling of nine simulated galaxies in halos of 
$10^{11.5}\Msun$. First, our analysis suffers from small number statistics. 
While we attempt to make full use of the nine galaxies by obtaining the mean
surface brightness profile from averaging all viewing angles (effectively 
creating a much larger sample for stacking), modeling more 
galaxies definitely helps the study of LAHs. More galaxies are also needed
to explore the dependence of LAHs on the properties of galaxies and their
environments \citep[e.g.,][]{Matsuda:2012aa}. 
Second, the mass of halos ($10^{11.5}\Msun$) considered in this 
work seems to be on the high end of LAE-hosting halos. The LAE halo masses
inferred from clustering analysis are typically $10^{11 \pm 1}\Msun$ 
\citep{Ouchi:2010aa}. Clearly it is necessary to investigate how the \lya and 
UV surface brightness profiles and their decomposition into cooling and 
star forming components vary with halo mass, and to make comparisons with
data especially as better observational constraints on the UV profile become 
available. As an example of the potential 
impact of halo mass, \citet{Rosdahl:2012aa} find that the extent 
of cooling radiation in the outer halo is dependent on halo mass, resulting 
from a positive correlation between the efficiency of cold streaming accretion 
and halo mass. Additional radiative transfer modeling of galaxies in lower 
mass halos comparing emission from the central LAE with star formation and 
cooling radiation in the outer halo will elucidate to what extent our results 
in this paper hold. 

We plan to carry out studies related to the halo mass and redshift
evolution of LAHs in order to obtain a better understanding of their origins and to learn 
more about the CGM and galaxy formation. 

\acknowledgments
We thank the anonymous referee for constructive comments and Mark Dijkstra for 
helpful discussions.
E.L., Z.Z., and R.S. are supported by NASA grant NNX14AC89G and NSF grant 
AST-1208891. The support and resources from the Center for High 
Performance Computing at the University of Utah are gratefully acknowledged. 
Computing resources were in part provided by the NASA High-End Computing 
(HEC) Program through the NASA Advanced Supercomputing (NAS) Division at 
Ames Research Center. R.C. is supported in part by grants NASA NNX11AI23G and 
NNX12AF91G. This work was supported by World Premier International Research
Center Initiative (WPI Initiative), MEXT, Japan, and KAKENHI
(23244025) Grantin-Aid for Scientific Research (A) through Japan
Society for the Promotion of Science (JSPS). 

\bibliography{bibliography}

\end{document}